# The microscopic structure of charge density waves in under-doped $YBa_2Cu_3O_{6.54}$ revealed by X-ray diffraction.

E. M. Forgan[1], E. Blackburn[1], A.T. Holmes[1], A. Briffa[1], J. Chang[2], L. Bouchenoire[3,4], S.D. Brown[3,4], Ruixing Liang[5], D. Bonn[5], W. N. Hardy[5], N. B. Christensen[6], M. v. Zimmermann[7], M. Hücker[8] and S.M. Hayden[9]

[1]School of Physics & Astronomy, University of Birmingham, Birmingham B15 2TT, UK

[2] Physik-Institut, Universität Zürich, Winterthurerstrasse 190, CH-8057 Zürich, Switzerland

[3]XMaS, European Synchrotron Radiation Facility, B. P. 220, F-38043 Grenoble Cedex, France

[4]Department of Physics, University of Liverpool, Liverpool, L69 3BX, UK

[5]Department of Physics & Astronomy, University of British Columbia, Vancouver, Canada

[6]Department of Physics, Technical University of Denmark, DK-2800 Kongens Lyngby, Denmark

[7]Deutsches Elektronen-Synchrotron DESY, 22603 Hamburg, Germany

[8]Condensed Matter Physics & Materials Science Department, Brookhaven National Laboratory, Upton, New York 11973, USA

[9]H. H. Wills Physics Laboratory, University of Bristol, Bristol, BS8 1TL, UK

**All underdoped high-temperature cuprate superconductors appear to exhibit charge density wave (CDW) order, but both the underlying symmetry breaking and the origin of the CDW remain unclear. We use X-ray diffraction to determine the microscopic structure of the CDW in an archetypical cuprate $YBa_2Cu_3O_{6.54}$ at its superconducting transition temperature $T_c \sim 60$ K. We find that the CDWs present in this material break the mirror symmetry of the $CuO_2$ bilayers. The ionic displacements in a CDW have two components: one *perpendicular* to the $CuO_2$**

**planes, and another parallel to these planes, which is out of phase with the first. The largest displacements are those of the planar oxygen atoms and are perpendicular to the $CuO_2$ planes. Our results allow many electronic properties of the underdoped cuprates to be understood. For instance, the CDW will lead to local variations in the doping (or electronic structure) giving an explicit explanation of the appearance of density-wave states with broken symmetry in scanning tunnelling microscopy (STM) and soft X-ray measurements.**

## Introduction

A charge density wave (CDW) is a periodic modulation of the electron density, associated with a periodic lattice distortion that may or may not be commensurate with the crystal lattice. The charge density modulation may be brought about by electron-phonon or electron-electron interactions[1]. It is now clear that the CDW state is a ubiquitous high-$T_c$ phenomenon, appearing in the underdoped region in both hole-[2-15] and electron-doped[16] cuprates, prior to the formation of the superconducting state, suggesting that the CDW is a characteristic instability of the $CuO_2$ plane. The CDW competes with superconductivity[2-4,17], and pressure-dependent data[18] suggest that if the CDW can be suppressed in $YBa_2Cu_3O_y$ (YBCO), then an enhanced $T_c$ occurs in the nominally underdoped region rather than at optimum doping. Experiments on YBCO using resonant soft X-ray scattering suggest that the CDW is associated with significant $d$-wave components for charges on the oxygen bonds around the Cu site[19,20], as proposed by Sachdev[21]. This conclusion is supported by scanning tunnelling microscopy (STM) observations of the surface of $Bi_2Sr_2CaCu_2O_{8+x}$ and $Ca_{2-x}Na_xCuO_2Cl_2$[22]. However, to understand the *generic* high $T_c$ CDW phenomenon, discovering the actual structure of the CDW is vital.

CDWs break the translation symmetry of the parent lattice, and have been observed by X-ray diffraction, and many other probes such as STM[22] and NMR[17,23]. Signatures of the Fermi surface reconstruction believed to be associated with this include quantum oscillation (QO) measurements[24], which show unexpectedly small Fermi surface pockets in an underdoped sample, and transport measurements, which indicate a change from hole carriers in the overdoped region to electron-like transport in the underdoped region[25]. To relate these observations to the CDW, its actual structure needs to be known. The studies of CDWs by X-ray diffraction in numerous cuprates have generally concentrated on determining the wavevector of the CDW and the temperature and magnetic field dependence of the order parameter and correlation lengths, and therefore have considered only a handful of diffraction satellites arising from the CDW. The only way to determine the structure unambiguously is by measuring the intensities of as many CDW satellites as possible. Here we make the first determination of the structure of the CDW in a bilayer cuprate. The material we have investigated is the well-studied material YBCO at a doping level where there is strong competition between superconductivity and the CDW, and the oxygen ordering in the crystal is most perfect.

**Experimental results**

To do this, we have used non-resonant X-ray diffraction to measure the intensities of all experimentally accessible CDW satellites near the ($h$, 0, $\ell$), (0, $k$, $\ell$) and ($h$, $h$, $\ell$) planes for both of the CDW modulation vectors, $\mathbf{q}_a = (\delta_a, 0, 0.5)$ and $\mathbf{q}_b = (0, \delta_b, 0.5)$. Throughout this paper, we express wavevectors in reciprocal space coordinates ($h$, $k$, $\ell$), where $\mathbf{Q}=(h\mathbf{a}^*+k\mathbf{b}^*+ \ell\mathbf{c}^*)$; $\mathbf{q}$ is used to denote the full wavevector of a CDW mode and $\delta$ its basal plane part. By collecting a comprehensive dataset, we deduce with great certainty the displacement patterns of the ions in the unit cell and hence the structure of the CDW in YBCO.

Non-resonant X-rays are primarily sensitive to the ionic displacements associated with a CDW, rather than changes in charge densities, although if one of these is present, so must the other[26]. The CDW order gives rise to very weak diffraction satellites at positions in reciprocal space $\mathbf{Q} = \boldsymbol{\tau} \pm \mathbf{q}$ around lattice Bragg peaks $\boldsymbol{\tau}$ which are at integer $h$, $k$, and $\ell$. The diffracted amplitude at wavevector $\mathbf{Q}$ due to an ion carrying a total of $N$ electrons displaced by small distance $\mathbf{u}$ is $\sim N\,\mathbf{Q}.\mathbf{u}$. Hence, the variation of the intensities with $\mathbf{Q}$ reflects the directions and magnitudes of the different ion displacements throughout the unit cell, and by observing intensities of CDW diffraction signals over a wide range of directions and values of **Q** we can determine the CDW structure. The full theory relating the CDW satellite intensities to the CDW structure is given in the Supplementary Information.

We may write the displacements $\mathbf{u}_j$, of the individual ions from their regular positions $\mathbf{r}_j^0$ as a sum of two terms, one of which is polarised along **c** ($\mathbf{u}_j^c$) and the other ($\mathbf{u}_j^a$ or $\mathbf{u}_j^b$) parallel to $\boldsymbol{\delta}$, with mirror symmetry about the relevant layer of the crystal.

$$\mathbf{r}_j = \mathbf{r}_j^0 + \mathbf{u}_j^c \cos(\boldsymbol{\delta}.\mathbf{r}_j^0 + \varphi) + \mathbf{u}_j^{a,b} \sin(\boldsymbol{\delta}.\mathbf{r}_j^0 + \varphi), \qquad (1)$$

Symmetry[27] requires that the $\mathbf{u}_j^c$ and $\mathbf{u}_j^{a,b}$ displacements are $\pi/2$ out of phase, as expressed in Eq. 1.

**The nature of the sample used.** Our experiment was carried out on an underdoped crystal with the ortho-II structure (meaning that the oxygen sites on alternate CuO chains are unoccupied). The crystal was the same as that used in Ref 7. Ortho-II was selected as it has been well studied by multiple techniques and the satellites associated with the oxygen-ordering have minimal overlap with the CDW satellites. Here, $\delta_a \sim 0.323$, $\delta_b \sim 0.328$ for our underdoped YBCO crystal with ortho-II structure.

Measurements were made at the superconducting $T_c$ of our sample (60 K), where the CDW intensity is a maximum in zero field[4]. CDW signals in high-$T_C$ cuprates are observed with basal-plane wavevectors along both **a** and **b** crystal directions. These modulations may be present in the crystal in *separate* domains having $\mathbf{q}_a$ or $\mathbf{q}_b$ modulation (a 1-**q** model); alternatively, *both* modulations could be present and superposed in the same region (a 2-**q** model). Intensity measurements at separate $\mathbf{q}_a$ and $\mathbf{q}_b$ do not interfere, so all qualitative features of the two CDW components that we may deduce from our results are independent of the 1- or 2-**q** state of the sample, which only affects numerical estimates of the absolute magnitudes of the displacements (by a factor of $\sqrt{2}$).

Fig. 1 (a-c) shows some typical scans through CDW diffraction satellites. They peak at half-integral values of ℓ (Fig. 1 (e)), and are extremely weak (~$10^{-7}$ of a typical crystal Bragg reflection). Therefore the satellites are measured above a relatively large background, but due to their known position and shape, their intensities can be found and spurious signals ignored (see Methods and Supplementary Information).

A compilation of some of the measured CDW intensities is displayed in Figure 2; the area of the red semicircles is proportional to the measured peak intensity.

**Ionic displacements obtained from the intensities.** Group theory indicates which incommensurately modulated displacement patterns or 'irreducible representations' (IRs) are consistent with the observed ordering wavevectors. There are four IRs for each ordering wavevector, labelled $A_1$, $A_2$, $A_3$, $A_4$ for $\mathbf{q}_a$ and similarly $B_1$ - $B_4$ for $\mathbf{q}_b$[27]. The even-numbered patterns have purely basal-plane transverse displacements, and are therefore incompatible with our observations of satellites close to both the **c*** axis and the basal plane[4]. The other IRs have longitudinal displacements in the basal plane parallel to $\delta$, combined with shear displacements parallel to the *c*–axis. Only the $A_1$ pattern for $\mathbf{q}_a$ (and $B_1$ for $\mathbf{q}_b$) are consistent with our data. These IRs have equal **c**-axis *shear* displacements in the two halves of the $CuO_2$ bilayer region, combined with basal plane compressive displacements (and hence charge density modulations) which are *equal and opposite* in the two halves of a bilayer. Thus these CDWs break the mirror symmetry of the bilayers. For these patterns, the CuO chain layer is a mirror plane of the CDW. For the patterns $A_3$ and $B_3$, the yttrium layer is instead a mirror plane of the CDW, so that the basal plane compressive displacements would be *equal* on the two sides of the bilayer.

Some ionic displacements in these IRs are zero by symmetry. This results in a detailed description of an IR that consists of 13 non-zero parameters representing displacement components of the 11 ions in the YBCO unit cell. In our model, we average over the half-occupied chain oxygen site in ortho-II YBCO, because we find no evidence for different responses in those cells having a full or empty CuO chain. (No CDW satellites were observed about $\tau + \frac{1}{2}\mathbf{a}^*$ positions). Our most complete dataset is for the $\mathbf{q}_b$ satellites. The $\mathbf{q}_a$ data are somewhat sparser and the results have larger errors, due to the tails of the peaks arising from the ortho-II oxygen-ordering which give large and rapidly varying backgrounds.

Models $A_1$ and $B_1$ always converged in a few iterations to a good fit and gave the same fitted values of displacements independent of the starting value of the parameters. In contrast, models $A_3$ and $B_3$ always gave poor fits (e.g. $\chi^2(B_3) > 10 \times \chi^2(B_1)$), whatever the starting values of the fitting parameters. Sample data and fits are shown in Fig. 2.

We therefore conclude that the IRs $A_1$ and $B_1$ are close to an accurate representation of the CDW. In Fig. 3, we represent the patterns of ionic displacements in a single unit cell as given by the data for both $\mathbf{q}_a$ and $\mathbf{q}_b$ modulations. The overall similarity of the two patterns is apparent. The spatial variation of the ionic displacements, shown in Fig. 4, is derived from the motifs in Fig. 3 by modulating the *c*-axis and basal plane displacements by $\cos(2\pi\delta_a x/a)$ and $\sin(2\pi\delta_a x/a)$ respectively for the $\mathbf{q}_a$ mode, and similarly for $\mathbf{q}_b$.

## Discussion

We have obtained an estimate of the absolute magnitude of ion displacements by comparing the satellite intensities with those of the Bragg peaks from the lattice. The fitted values of the ionic displacements are given in the Supplementary Information. Earlier measurements found differences in the magnitudes of the X-ray signals from the $\mathbf{q}_a$ and $\mathbf{q}_b$ modes[6,7] that suggested that the CDW in o-II YBCO might be essentially single-$\mathbf{q}$, and dominated by the $\mathbf{q}_b$ mode. The results presented here indicate that this is *not* the case; the two modes have similar displacement amplitudes, but the value of their ratio depends on which ion is chosen to make the comparison. For instance, if we consider the motion of the $CuO_2$ plane oxygens as key, we find that the *relative c*-motion of these oxygen ions is essentially identical for the two modes: in both cases, the amplitude is ~ 4-5 × $10^{-3}$ Å. Our fits do show differences in the heavy ion displacements, and even if these are small, they can make noticeable contributions to the X-ray signals because these ions carry many core electrons. Only this complete survey, rather than measurements of the intensities of a few $\mathbf{q}_a$ and $\mathbf{q}_b$ satellites, can reveal the similarities and differences between the two CDW modes.

The deduced ionic displacements are maximal near the $CuO_2$ planes and weak near the CuO chains; this is in agreement with the observed competition between the CDW and superconductivity[2-4,17]. Surprisingly, the largest amplitudes are out-of-plane *shear* rather than compression of the $CuO_2$ planes, so that the CDW is not *purely* a separation of charge as commonly assumed. It may be that the lattice is deforming in this way because shear deformations cost less elastic energy than compressive ones. There are also CDW-modulated charges associated with the small longitudinal displacements in the two halves of a bilayer, but they are equal and opposite. This would be favoured by Coulomb effects within a bilayer.

We note the similarity of some of the displacements to a soft phonon observed in optimally doped YBCO[28]. However, in that mode, the *c*-motion is in *anti*phase for the two halves of the bilayer. Buckling of the $CuO_2$ planes is also seen in 214 compounds[29], where it mainly consists of tilts of rigid Cu-O octahedra. Here however, the displacements in the $CuO_2$ layers are clearly inconsistent with tilts of a rigid arrangement of ions.

**Comparison with STM and polarised x-rays.** We draw attention to the up/down "butterfly" nature of the displacements of the four oxygens around a Cu in the bilayers, which is seen for both $\mathbf{q}_a$ and $\mathbf{q}_b$ modes. The two oxygens in the δ-direction around a copper are displaced in the *same* direction as the Cu along **c**, but the other perpendicular pair is displaced *oppositely* (see Fig. 3). To an STM[22] this could appear as a '*d*-charge density' on the oxygens, since *c*-axis motion of an oxygen – relative to the yttrium and/or to the crystal surface would alter its local doping and electronic state. We note that the STM measurements are analysed in such a way as to emphasise the electronic states, rather than the positions of atoms. In Fig. 5(a), we show qualitatively what the effect on the local doping of the oxygen ions might be by assuming that the change is proportional the displacement along **c**. The pattern produced has the same symmetry as that observed by STM[22] in $Ca_{2-x}Na_xCuO_2Cl_2$ [see Fig. 5(b)]. STM and azimuthal angle-dependent resonant x-ray (RXS) studies[19,20] of the charge order have been analysed in terms of modulated states with local symmetry of three types with respect to a planar copper site: equal density on the copper atoms (*s* symmetry); equal density on the neighbouring oxygen atoms (*s'* symmetry); opposite-sign density on the neighbouring $O_x$ and $O_y$ sites (*d*-symmetry). Our measured copper and oxygen displacements (see SI) can provide an explanation for the relative proportions of these components. In agreement with the STM and RXS, we find that the *d*-wave component is dominant.

## Outlook and Perspectives

These results carry several important messages. Firstly, they show that a strictly "planar" account of high-$T_c$ phenomena may miss important aspects of the physics, and that the 3$^{rd}$ dimension and crystal lattice effects cannot be ignored. In our experiments we have observed a charge density wave with a strong *shear* (**c**-axis) component. The "butterfly" pattern of oxygen shear displacements around the planar copper ions can *simulate* a *d*-charge density on the oxygens. It will be very interesting to repeat these X-ray measurements on other underdoped high-$T_c$ compounds to establish the generality (or otherwise) of these results, and to relate these results to the changes in the CDW that occur at high fields where quantum oscillation measurements are performed. Ultrasonic measurements[30] show that changes occur at about 18 T. Very recent measurements in pulsed field[31] in an YBCO sample with o-VIII oxygen ordering show that longer-range order with the same value of $\delta_b$ emerges at high field. This is clearly related to our zero-field structure, and leads to interesting questions[32] about the Fermi surface reconstruction at low and high fields. It is clear that antiferromagnetic order, the CDW, pseudogap and superconductivity are all intertwined, since they all remove electron states near the antinodal regions of the Fermi surface. It appears that there is a quantum critical point underlying the superconducting dome; we trust that our results will help to achieve an over-arching theory relating the relationship of all these phenomena to high-$T_c$ superconductivity.

## Methods

To obtain sufficient data required the flexibility of a 4-circle diffractometer, which is provided at the XMaS beamline at the ESRF, Grenoble[33]. The sample was mounted in a closed-cycle cryostat and all measurements were carried out in zero magnetic field in reflection from the flat *c*-face of the crystal (of area ~ 2 × 2 mm$^2$) at an X-ray energy of 14 keV. This gives a penetration depth of 25 microns into the sample, so the results are not dominated by surface effects. For CDW intensity measurements, the sample temperature was controlled at $T_c$ to maximise the signal, and it was taken to 150 K to check for spurious signals which did not go to zero. The diffractometer angles were set so that the incoming and detected beams were close to the same angle to the *c*-face of the crystal, which allowed correction for sample absorption, as described in the Supplementary Information. CDW intensity measurements were carried out near the (*h*, 0, ℓ), (0, *k*, ℓ) and (*h*, *h*, ℓ) planes of reciprocal space over as wide a range of *h*, *k* and ℓ allowed by the maximum scattering angle, and the avoidance of grazing incidence at low ℓ. CDW peaks were scanned parallel to $\delta$, through positions of the form **Q** = **τ**±**q**. The intensities of the CDW peaks were established by fitting each scan with a Gaussian of fixed width, with a smoothly-varying cubic polynomial background. By examination of 150 K measurements, or by the $\chi^2$ of the fit, spurious peaks were removed from the list of measured satellites. As shown in Fig. 2, the finite range of the CDW order results in satellites that are broad, particularly in the **c**\* direction. However, all the intensity of any satellite is confined to a single Brillouin zone, allowing it to be integrated over reciprocal space. The resulting list of intensities, weighted by their errors, was fitted to our CDW models by varying the ionic displacements {$\mathbf{u}_j$} to minimise $\chi^2$. Further details are in the Supplementary Information.

## Acknowledgements

We thank Martin Long and Radu Coldea for very helpful discussions and the UK EPSRC for funding under grant numbers EP/J016977/1 (EB, EMF & ATH) and EP/J015423/1 (SMH). JC wishes to thank the Swiss National Science Foundation for support. NBC was supported by the Danish Agency for Science, Technology and Innovation under DANSCATT. Sample preparation was funded through NSERC and the Canadian Institute for Advanced Research. Work at Brookhaven is supported by the Office of Basic Energy Sciences (BES), Division of Materials Science and Engineering, U.S. Department of Energy (DOE), under Contract No. DE-AC02-98CH10886. We thank U. Ruett and D. Robinson for invaluable assistance with complementary higher energy measurements performed at P07, DESY & 6-ID-D, APS. Use of the Advanced Photon Source, an Office of Science User Facility operated for the U.S. DOE Office of Science by Argonne National Laboratory, was supported by the U.S. DOE under Contract No. DE-AC02-06CH11357.

## Author Contributions

EB, JC, EMF, SMH & ATH carried out the experiments with important input from beamline scientists LB & SDB. Analysis of results was carried out by EB, JC, EMF, SMH & ATH; AB created the program that fitted the intensities to the ionic displacements. NBC, MH & MvZ contributed additional measurements at high energy. DB, WH & RL supplied samples. All authors contributed to the manuscript.

# References

1. Zhu, X., Cao, Y., Zhang, J., Plummer, E. W. and Guo. J. Classification of charge density waves based on their nature. *Proc. Nat. Acad. Sci.* **112**, 2367-2371 (2015).

2. Ghiringhelli, G. *et al*. Long-Range Incommensurate Charge Fluctuations in (Y,Nd)Ba$_2$Cu$_3$O$_{6+x}$. *Science* **337**, 821-825 (2012).

3. Achkar, A.J. *et al.*, Distinct Charge Orders in the Planes and Chains of Ortho-III-Ordered YBa$_2$Cu$_3$O$_{6+\delta}$ Superconductors Identified by Resonant Elastic X-ray Scattering. *Phys. Rev. Lett.* **109**, 167001 (2012).

4. Chang, J. *et al.* Direct observation of competition between superconductivity and charge density wave order in YBa$_2$Cu$_3$O$_{6.67}$. *Nature Physics* **8**, 871-876 (2012).

5. Blanco-Canosa, S. *et al.* Momentum-Dependent Charge Correlations in YBa$_2$Cu$_3$O$_{6+x}$ Superconductors Probed by Resonant X-ray Scattering: Evidence for Three Competing Phases. *Phys. Rev. Lett.* **110**, 187001 (2013).

6. Blanco-Canosa, S. *et al.* Resonant X-ray scattering study of charge-density wave correlations in YBa$_2$Cu$_3$O$_{6+x}$. *Phys. Rev. B* **90**, 054513 (2014).

7. Blackburn, E. *et al.* X-ray Diffraction Observations of a Charge-Density-Wave Order in Superconducting Ortho-II YBa$_2$Cu$_3$O$_{6.54}$ Single Crystals in Zero Magnetic Field. *Phys. Rev. Lett.* **110**, 137004 (2013).

8. Huecker, M *et al.* Competing charge, spin, and superconducting orders in underdoped YBaCuO$_y$. *Phys. Rev. B* **90**, 054514 (2014).

9. Comin, R. *et al.* Charge Order Driven by Fermi-Arc Instability in Bi$_2$Sr$_{2-x}$La$_x$CuO$_{6+d}$. *Science* **343**, 390, (2014).

10. da Silva Neto *et al.* Ubiquitous Interplay Between Charge Ordering and High-Temperature Superconductivity in Cuprates. *Science* **343**, 393-396 (2014).

11. Hashimoto, M. *et al.* Direct observation of bulk charge modulations in optimally doped Bi$_{1.5}$Pb$_{0.6}$Sr$_{1.54}$CaCu$_2$O$_{8+d}$. *Phys. Rev. B* **89**, 220511 (2014).

12. Thampy, V. *et al.* Rotated stripe order and its competition with superconductivity in La$_{1.88}$Sr$_{0.12}$CuO$_4$. *Phys. Rev. B* **90**, 100510 (2014).

13. Tabis, W. *et al*. Charge order and its connection with Fermi-liquid charge transport in a pristine high-$T_c$ cuprate. *Nature Communications* **5**, 5875 (2014).

14. Christensen, N. B. *et al*. Bulk charge stripe order competing with superconductivity in La$_{2-x}$Sr$_x$CuO$_4$ ($x$=0.12). Preprint at http://arxiv.org/abs/1404.3192 (2014).

15. Croft, T. P., Lester, C., Senn, M. S., Bombardi, A. and Hayden, S. M. Charge density wave fluctuations in La$_{2-x}$Sr$_x$CuO$_4$ and their competition with superconductivity. *Phys. Rev. B* **89**, 224513 (2014).

16. E.H. da Silva Neto *et al.*, Charge ordering in the electron-doped superconductor Nd$_{2-x}$Ce$_x$CuO$_4$, *Science* **347**, 282-285 (2015).

17. Wu, T. *et al.*, Magnetic-field-induced charge-stripe order in the high temperature superconductor YBa$_2$Cu$_3$O$_y$. *Nature* **477**, 191–194 (2011).

18. O. Cyr-Choinière *et al.*, Suppression of charge order by pressure in the cuprate superconductor YBa$_2$Cu$_3$O$_y$ : restoring the full superconducting dome. Preprint at http://arxiv.org/abs/1503.02033 (2015).

19. Comin, R. *et al.* Symmetry of charge order in cuprates. *Nature Materials,* Advance Online Publication, doi: 10.1038/NMAT4295 (2015).

20. Achkar, A. J. *et al.* Orbital symmetry of charge density wave order in La$_{1.875}$Ba$_{0.125}$CuO$_4$ and YBa$_2$Cu$_3$O$_{6.67}$. Preprint at http://arxiv.org/abs/1409.6787 (2014).

21. Sachdev, S. and La Placa, R. Bond order in two-dimensional metals with antiferromagnetic exchange interactions. *Phys. Rev Lett.* **111**, 027202 (2013).

22. Fujita, K. *et al.* Simultaneous Transitions in Cuprate Momentum-Space Topology and Electronic Symmetry Breaking. *Science* **344**, 612-616 (2014).

23. Wu, T. *et al.* Incipient charge order observed by NMR in the normal state of $YBa_2Cu_3O_y$. *Nature Communications* **6**, 6438 (2015).

24. Doiron-Leyraud, N. *et al.* Quantum oscillations and the Fermi surface in an underdoped high-$T_c$ superconductor. *Nature* **447,** 565-568 (2007).

25. Chang, J. *et al.* Nernst and Seebeck coefficients of the cuprate superconductor $YBa_2Cu_3O_{6.67}$: a study of Fermi surface reconstruction. *Phys. Rev. Lett.* **104**, 057005 (2010).

26. Abbamonte, P., Charge modulations versus strain waves in resonant X-ray scattering. *Phys. Rev. B* **74**, 195113 (2006).

27. Campbell, B. J., Stokes, H. T., Tanner, D. E. and Hatch, D. M. ISODISPLACE: An Internet Tool for Exploring Structural Distortions. *J. Appl. Cryst.* **39**, 607-614 (2006).

28. Raichle, M. *et al.*, Highly Anisotropic Anomaly in the Dispersion of the Copper-Oxygen Bond-Bending Phonon in Superconducting $YBa_2Cu_3O_7$ from Inelastic Neutron Scattering. *Phys. Rev. Lett.* **107**, 177004 (2011).

29. Tranquada, J. M. Stripes and superconductivity in cuprates. *Physica B: Condensed Matter* **407**, 1771-1774 (2012*)*.

30. LeBoeuf, D. *et al.*, Thermodynamic phase diagram of static charge order in underdoped $YBa_2Cu_3O_y$. *Nature Physics* **9**, 79–83 (2013).

31. Gerber, S. *et al*., Three-Dimensional Charge Density Wave Order in $YBa_2Cu_3O_{6.67}$ at High Magnetic Fields. Preprint at http://arxiv.org/abs/1506.07910 (2015).

32. Briffa, A., Forgan, E. M., Blackburn, E., Yelland, E. A., Hayden, S. M. and Long, M. W. Can the Fermi Surface Reconstruction and Quantum Oscillations in Under-doped $YBa_2Cu_3O_{6.54}$ be explained by Charge Density Waves? *in preparation* (2015).

33. Brown S. D. *et al.*, *The XMaS beamline at ESRF: instrumental developments and high resolution diffraction studies. J. Synch. Rad.* **8**, 1172-1181 (2001); http://www.xmas.ac.uk.

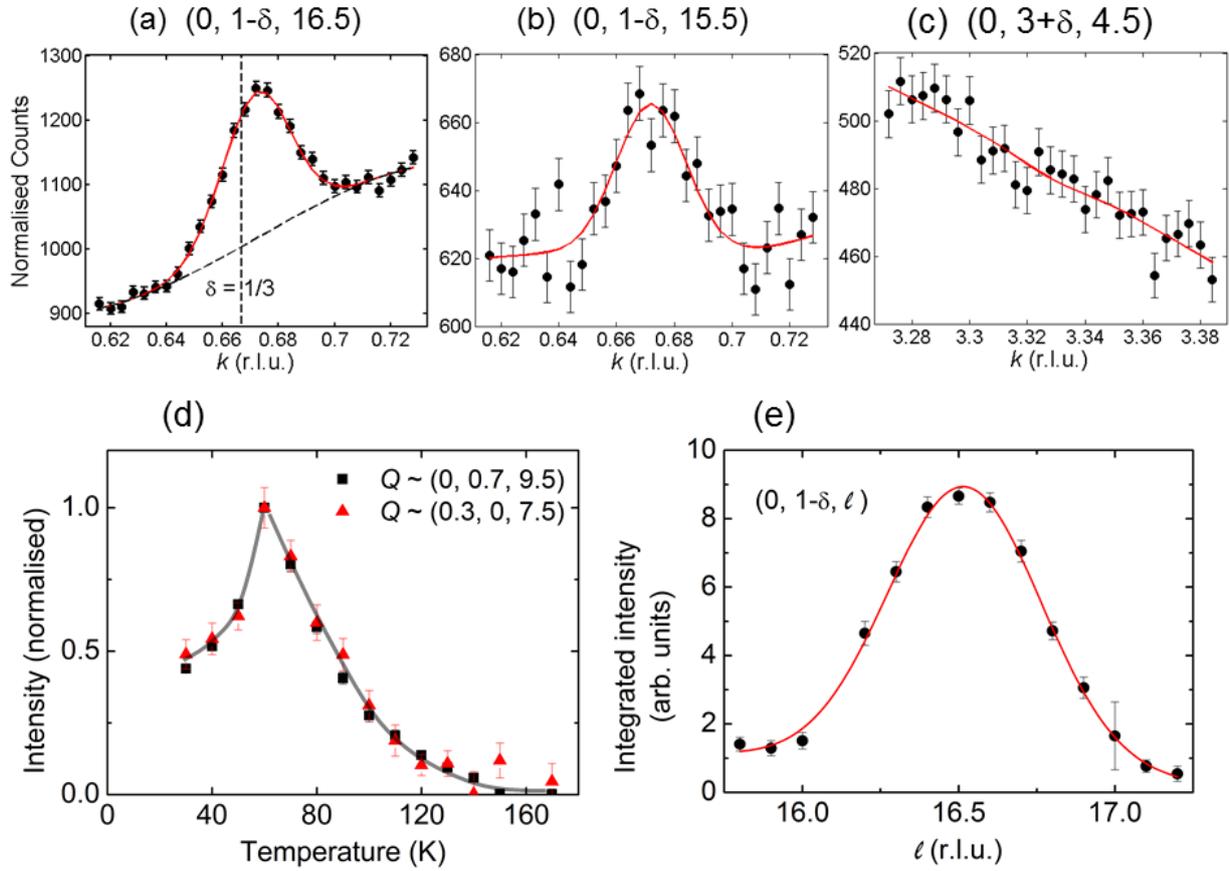

**Figure 1: Typical observations of CDW satellites at 60 K and their temperature-dependence**

(a)-(c) are obtained from the CDW with modulation vector $\mathbf{q}_b$. The counts are normalised so that they are approximately per second, measured over 10 s per point, plotted versus wavevector along the $\mathbf{b}^*$ direction, labelled $k$. (a) shows a strong satellite, along with the fit line which gives the intensity as the area under the peak. The CDW is clearly centred at an incommensurate position ($\delta_b$ ~0.328), although the value ⅓ lies within the peak. (b) shows a weaker peak, and (c) is taken at a position where the CDW signal is unobservably small, and the fitted area of the peak is controlled by Poisson errors. (d) The intensities of CDW satellites for both $\mathbf{q}_a$ ($\delta_a$ ~0.323) and $\mathbf{q}_b$ ($\delta_b$ ~0.328) modes, normalised to their intensities at $T_c$, are plotted versus temperature; these track each other within errors. (e) The integrated intensity of the satellite (a) is plotted versus $\ell$. The width in $\ell$, which reflects the finite $\mathbf{c}$-axis coherence of the CDW, is much larger than the instrument resolution. Since it is a property of the CDW, it is the same for all satellites.

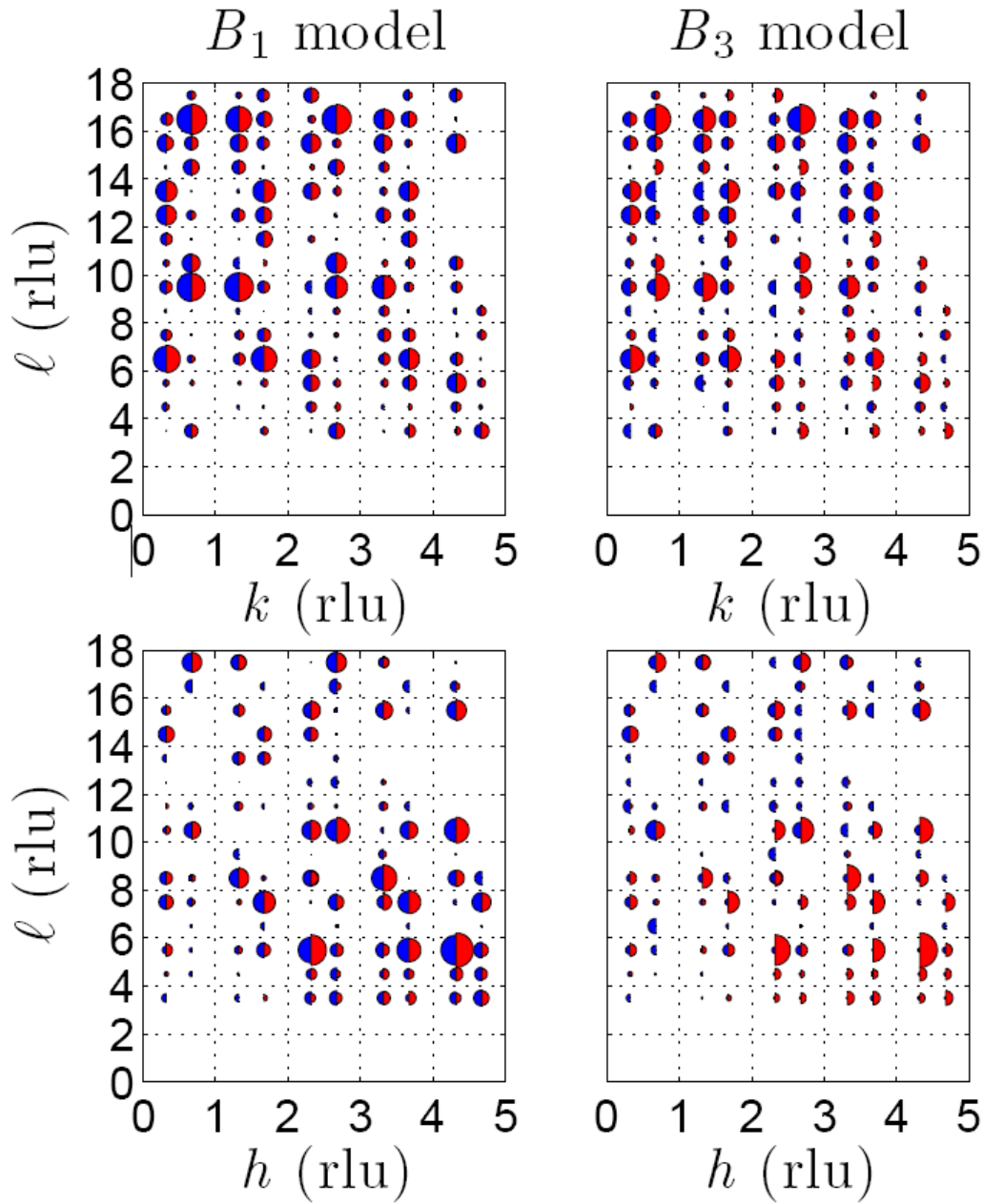

**Figure 2: Sample data compared with the fits to the two possible models.**

The upper panels show a map of the satellite intensities associated with CDW modulation $\mathbf{q}_b$ as measured in the $(0, k, \ell)$ plane of reciprocal space. The lower panels show a map of the satellite intensities associated with CDW modulation $\mathbf{q}_a$ as measured in the $(h, 0, \ell)$ plane of reciprocal space. The measured intensities are proportional to the areas of the red semicircles on the right of each **Q**-point. The blue semicircles show the results of two different fits (to all measured data, not just that shown in the Figure), to models described in the main text. Blank spaces indicate inaccessible regions or where a spurious signal prevented measurements of the CDW order. Maps of all other intensities that were measured are given in Supplementary Information.

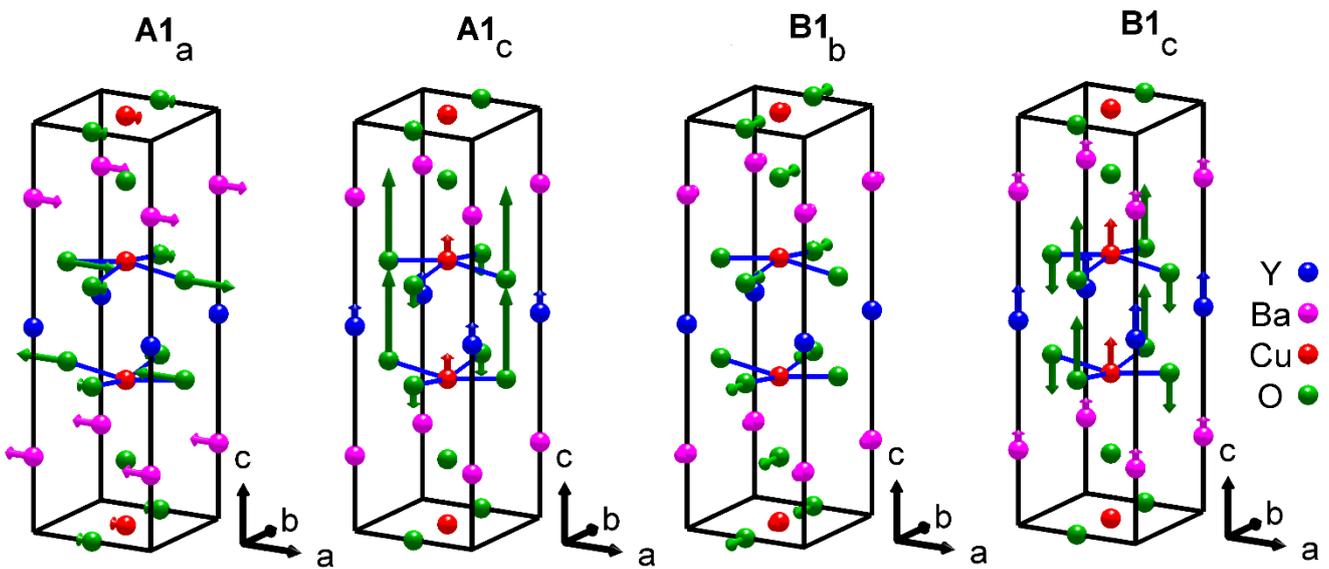

**Figure 3: Representation of the CDW ionic displacement motifs (exaggerated) for an unmodulated unit cell**

These are shown separately for the $\mathbf{q}_a$ and $\mathbf{q}_b$ modes of the CDW. The basal plane and *c*-axis displacements have a $\pi/2$ phase difference and hence are shown in separate unit cells. The next crystal unit cells in the *c*-direction would be in antiphase with those shown here. The oxygen sites in the CuO chains represented here are half-occupied in our sample.

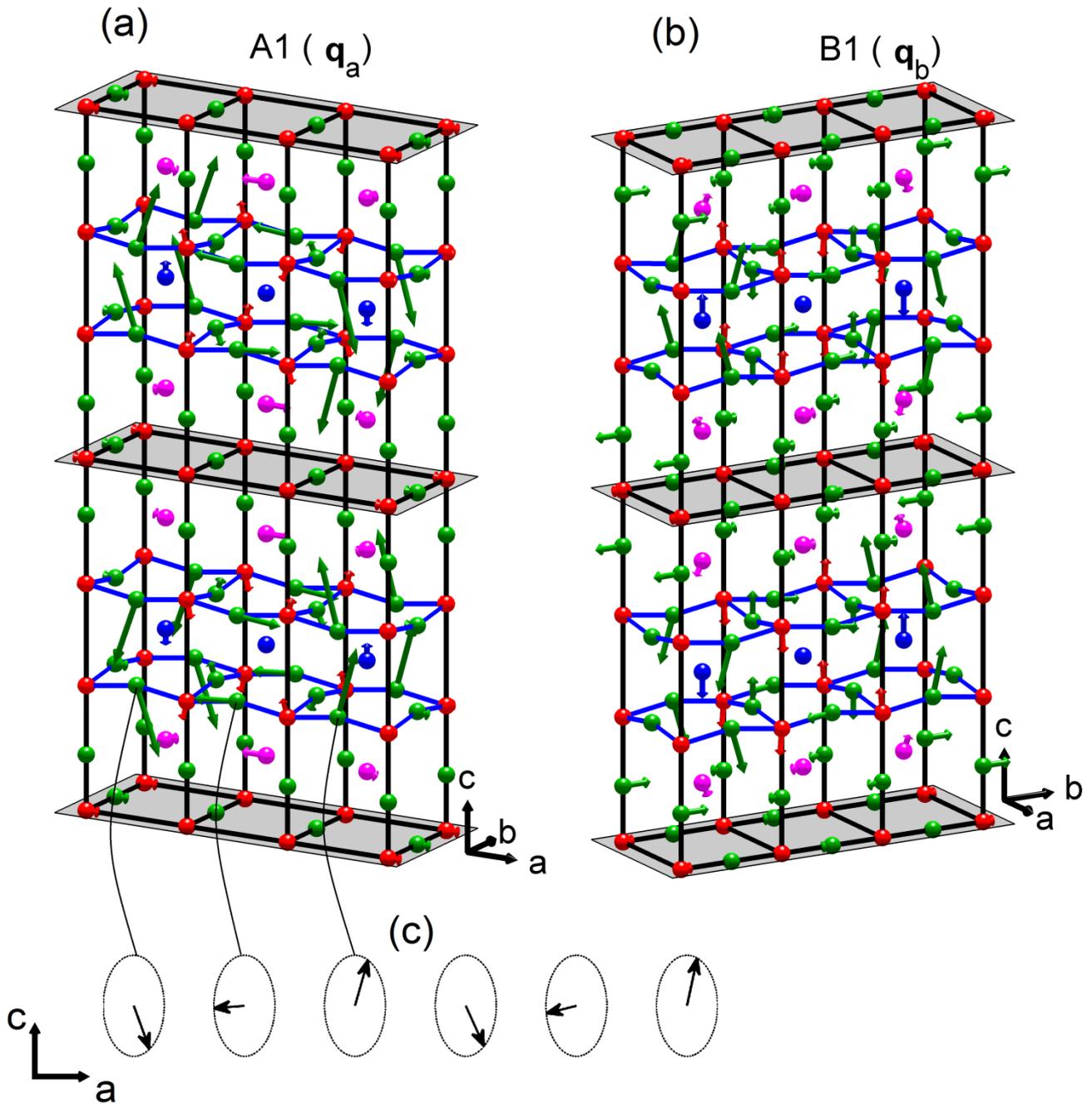

**Figure 4: Representation of modulated ionic displacements (exaggerated) for the CDW modes.**

The spatial variation of the ionic displacements for the $q_a$ and $q_b$ CDW modes is shown. The shaded planes passing through the CuO chain layers are the mirror planes of the CDWs. If the structure of the CDW is 1-**q**, these displacement patterns would be located in different regions of the crystal. If 2-**q**, the total displacement of the ions in the crystal would be the sum of those associated with the $q_a$ and $q_b$ modulation vectors. (c) The displacement of any particular ion in a CDW lies on an ellipse: we give an example for an oxygen in the lowest $CuO_2$ plane of panel (a).

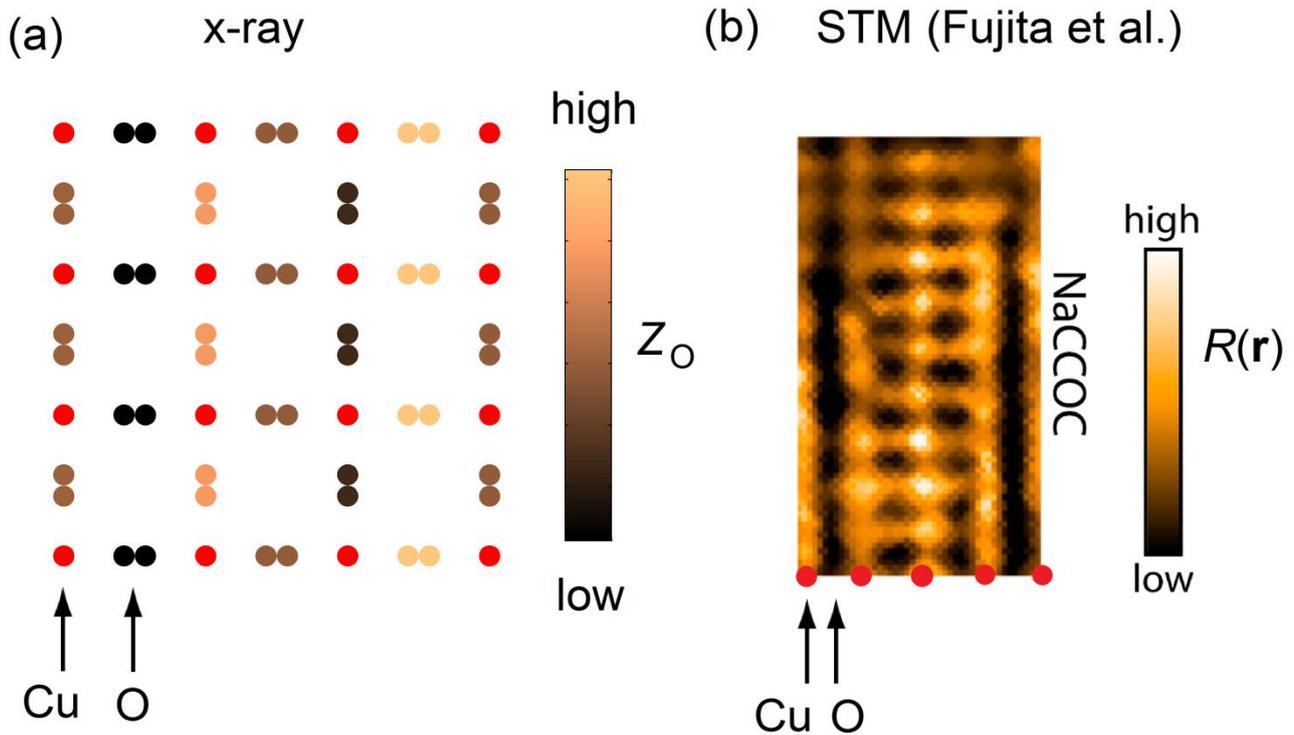

**Figure 5: Comparison of the bilayer oxygen height with a STM R(r) image.** (a) A representation of the spatial variation of the $z$ co-ordinate of the oxygen bilayer atoms shown for one ($\mathbf{q}_a$) of two modulation directions present in the crystal. (b) STM $R(\mathbf{r})$ image, where $R(\mathbf{r}) = I(\mathbf{r},E)/-I(\mathbf{r},-E)$ of the lightly doped cuprate $Ca_{2-x}Na_xCuO_2Cl_2$ (NaCCOC) (reproduced with permission from Ref. 22). The $R(\mathbf{r})$-image is used to highlight spatial variation of doping or electronic structure. It has the same symmetry as a "bond $d$-density wave" along the $\delta_a$ direction. Note that NaCCOC has a repeat period of approximately four unit cells, which is longer than that in YBCO.